# Implementation of Convolutional Neural Network Architecture on 3D Multiparametric Magnetic Resonance Imaging for Prostate Cancer Diagnosis

Ping-Chang Lin[1], Teodora Szasz[1], and Hakizumwami B. Runesha[1]

## Abstract

Prostate cancer is one of the most common causes of cancer deaths in men. There is a growing demand for noninvasively and accurately diagnostic methods that facilitate the current standard prostate cancer risk assessment in clinical practice. Still, developing computer-aided classification tools in prostate cancer diagnostics from multiparametric magnetic resonance images continues to be a challenge. In this work, we propose a novel deep learning approach for automatic classification of prostate lesions in the corresponding magnetic resonance images by constructing a two-stage multimodal multi-stream convolutional neural network (CNN)-based architecture framework. Without implementing sophisticated image preprocessing steps or third-party software, our framework achieved the classification performance with the area under a Receiver Operating Characteristic (ROC) curve value of 0.87. The result outperformed most of the submitted methods and shared the highest value reported by the PROSTATEx Challenge organizer. Our proposed CNN-based framework reflects the potential of assisting medical image interpretation in prostate cancer and reducing unnecessary biopsies.

## Introduction

Prostate cancer (PCa), the second and fifth leading causes of death from cancer in men in the United States and worldwide, can mostly be detected by screening (Hoffman, 2011). The current gold-standard prostate cancer diagnosis procedure is serum prostate-specific antigen level screening, followed by a transrectal ultrasonography-guided prostate biopsy (Kasivisvanathan *et al*., 2018; Stabile *et al*., 2020). However, the current diagnostic pathway also results in clinically significant levels of misdiagnosis and overdiagnosis. Given the positive correlation between tumorous volume and tumor grade (Fütterer *et al*., 2015), noninvasive multiparametric MRI (mpMRI) has become essential for improving prostate cancer staging and mitigating diagnostic errors since it was introduced in the early 1980s. Moreover, the Prostate Imaging Reporting and Data System (PI-RADS) was developed to lubricate effective communication between radiologists and urologists, increase the agreement between readers, and lessen the skill and experience gaps between the radiologists and the medical centers (Turkbey *et al*., 2019). Still, the limited inter-rater and intra-rater agreement, reflecting high subjectivity, remains unsolved even after the introduction of PI-RADS v2 (Meyer *et al*., 2019).

Computer-aided detection or diagnosis (CAD) exploits machine learning algorithms to circumvent the variation attributed to inter-observer reliability in prostate cancer diagnosis. Chan *et al*. proposed a multichannel classifier system that utilized co-occurrence matrix and discrete cosine transform to extract texture features from T2-weighted (T2w), T2 map, apparent diffusion coefficient (ADC), and proton density images (Chan *et al*., 2003). The texture features were then fed into the support vector machines (SVMs) or Fisher linear discriminant algorithms to create a map illustrating malignancy likelihoods in a prostate gland's peripheral zone. In addition to adopting the T2 map and ADC images, Ozer *et al*. embraced $K_{ep}$ maps derived from dynamic contrast-enhanced (DCE) MRI for automated prostate cancer

1. P-C. Lin (e-mail: pcclin@uchicago.edu), T. Szasz (e-mail: tszasz@uchicago.edu), and H. B. Runesha (e-mail: runesha@uchicago.edu) are with the Research Computing Center at the University of Chicago, Chicago, IL 60637 USA. Corresponding author: P-C. Lin segmentation (Ozer *et al.*, 2010). The threshold selecting scheme was used to tune the automated segmentation methods, including SVMs, relevance vector machines, and unsupervised fuzzy Markov random fields, for optimizing performance in the desired metrics. In the reference reported by Litjens *et al.*, the authors proposed an approach consisting of the detection and diagnosis stages (Litjens *et al.*, 2014). The prostate zone segmentation was performed on the T2w images at the detection stage, followed by image voxel feature extractions and voxel classification to generate a prostate malignancy likelihood map. After the highest probabilities in the averaged lesion size regions were retrieved from prostate MRI images by exploiting local-maximum detection, the diagnosis stage performed a region segmentation to merge the candidate regions that were further classified to present the associated cancer likelihoods. In addition, several similar studies aimed at prototyping CAD systems, which mainly employed a two-stage classification procedure consisting of prostate gland segmentation, voxel-wise feature extraction and classification to generate probability maps of malignancy based on mpMRI inputs, and candidate lesion combination and localization for final classification (Artan *et al.*, 2010; Liu and Yetik, 2011; Shah *et al.*, 2012; Wang *et al.*, 2014).

Several CAD systems implementing conventional machine learning algorithms have demonstrated the facilitation of medical image interpretation in clinical research (Giger, 2018; Wernick *et al.*, 2010). Still, crafting image features depends heavily on domain expertise, and finding and extracting appropriate image features is challenging for a specific medical image recognition problem. Moreover, the extensive variation in subject image data renders the detection or diagnosis system unreliable (Wang *et al.*, 2017). With promising results in the computer vision field, researchers have explored neural networks that aimed at recognizing intricate patterns in the cross-sectional radiographic images in the past few years (Hosny *et al.*, 2018). Among the neural network models adopted and applied to medical image analysis problems, convolutional neural network (CNN)-based approaches to prostate magnetic resonance imaging (MRI) have revealed a progressive increase *(Khalvati et al., 2019; Wang et al., 2018; Zabihollahy et al.*, 2020). A recent study proposed a double-modal fusion framework that the ADC and T2w images of MRI were implemented to train the respective CNN architecture streams in parallel (Yang *et al.*, 2017). The last convolutional layer in each of the CNN streams generated a pixel-wise cancerous response map. Furthermore, the PCa-relevant feature vectors obtained from the CNN architectures were considered the inputs of an SVM classifier during the training process. Upon the adaptive thresholding and non-maximum suppression, the SVM classifier's outcome was used to screen the locations of malignant prostate lesions. Liu *et al.* introduced a VGGNet inspired architecture for prostate lesion classification (Liu *et al.*, 2017). The multiparametric MRI dataset, adopted from the PROSTATEx challenge of possessing diffusion-weighted imaging (DWI), T2w, ADC, and $K^{trans}$ images for each case, was implemented to train the CNN-based model. The image data were preprocessed to pose the same isotropic resolution and refine the lesion centers, followed by the image data augmentation, including in-plane rotation, multiple orientation slicing, random shearing, and translation. The model inputs, referring to the image patches of the 32 pixels x 32 pixels in a region of interest, were extracted surrounding the lesion center; the classification outcome was determined by the average of the predictions resulting from different slicing orientations. Yoo *et al.* conducted a two-level deep learning pipeline containing five ResNet architectures at the slice level and a combination of first-order statistical feature extraction, decision tree-based feature selection, and random forest classification at the patient level for detection of clinically significant prostate lesions (Yoo *et al.*, 2019). First, the standardized DWI slices were fed into the five individual CNN models to generate the slice-level classification probabilities, from which the first-order statistical features were extracted for each case. The extracted statistical features were further filtered through the decision tree algorithm and subsequently categorized either to the clinical significance group or to the clinical insignificance group by conducting a Random Forest classifier. Another recent study published by Cao *et al.* exploited a multi-class CNN architecture to classify PCa lesions pixel-wisely from T2w and ADC images (Cao *et al.*, 2019). Once processing image registration and intensity normalization, the corresponding T2w and ADC image slices were fed into the DeepLab-based architecture for joint pixel-level prostate cancer and fine-grained Gleason score predictions.

In the current work, we propose a two-stage framework of a simple data preparation pipeline and construct a novel CNN-based neural network architecture to classify clinically significant prostate cancers for the given detected lesions. We also studied the prostate MRI dataset adopted from the PROSTATEx challenge in advance to strategize the data preprocessing pipeline. Consequently, we spatially resampled the images to unify the image size and resolution across the different MRI modalities in all the study cases, followed by processing the MRI modality-based featurewise standardization of all images. Instead of applying a third-party software or a sophisticated algorithm to perform image registration, we aligned the cropped prostate images in all the selected MRI modalities per the clinical finding labels' coordinates. For every study, the aligned images in the different types of MRI modalities were stacked to form a multichannel 3D image used to generate image patches of four different patch sizes. Upon the sizes of image patches, we constructed the corresponding quadruple-modal architecture of using DenseNet as the network backbone to make the respective predictions, which were further assembled into a two-level meta-learning network to improve the classification performance and to reduce the likelihood of prediction bias. We also adopted a novel loss function named Focal Loss that dynamically scales the cross-entropy loss to efficiently deal with class imbalance (Lin *et al.*, 2020, 2017). In our knowledge this is the first time that the multi-stream multimodal architecture of different image patch sizes integrates with Focal loss function to emphasize harder tamed samples during the model training and the second stacked generalization framework assemble the multiply model streams to mitigate the impact of overfitting for prostate cancer MRI classification.

## Methods

### Datasets

In this work, a prostate MRI dataset was gathered from the retrospective studies performed at the Prostate MR Reference Center in the Radboud University Medical Centre (Nijmegen, the Netherlands) (Armato *et al.*, 2018). The dataset consists of 349 studies (346 patients), in which 204 subjects were in the training cohort, containing the 330 lesion labels, of which 76 are clinically significant, and the remaining 142 subjects of 208 lesion labels with unknown clinical significances were in the test cohort. Every study includes at least T2w, dynamic contrast-enhanced (DCE), and DWI acquired on either of the two different Siemens 3T MRI scanners. Apparent Diffusion Coefficient (ADC) maps and $K^{trans}$ were calculated from DWI and DCE images, respectively. Upon the PROSTATEx organizer, the biopsy Gleason score of 7 or higher was considered a clinical significant finding, while findings of PI-RADS score 2 were neither considered clinically significant nor biopsied. The loci of these findings were coordinated in the prostate MRI images regardless of clinical significances.

### Data Preprocessing

### Image Scaling, Cropping and Standardizing

We observed variations in spatial resolution and field of view (FOV) across the MRI modalities, including T2w, ADC, DWI, and $K^{trans}$, in the same prostate cancer studies within the dataset. In addition, a similar scenario occurs when comparing the different studies in the same dataset. Therefore, rescaling and cropping the respective images across the different MRI modalities and subject studies are expected to unify the spatial resolution and FOV, which becomes a prerequisite for training a convolutional neuronal network. After excluding 12 subjects that contain different image slice thickness, mismatching finding coordinates across the different MRI modalities in the same study cases, or relatively low image quality from the training cohort (Liu *et al.*, 2017), the remaining 192 subjects contained the cropped images in a center sub-volume of 320 pixels × 320 pixels in size, with a resolution of 0.5 mm/pixel × 0.5 mm/pixel and a slice thickness of 3 mm, which is the setting for the majority of T2w image slices. Subjects in the test cohort were processed in the same manner. The mean and standard deviation of the cropped image pixel values per MRI modality were calculated in the training cohort before standardizing all the images accordingly.

### Image Patch Retrieval

The cropped images retrieved from the four different MR modalities in each study were translationally aligned per the clinical findings' coordinates rather than sophisticated image registration algorithms. The aligned, cropped prostate MRI images, excluding $K^{trans}$, were then concatenated into the 3-channel composite images. These 3-channel image slices evaluating prostatic sections from the base to the apex were further extracted to generate the 3D 3-channel image patches (Fig. 1A). In addition, the corresponding $K^{trans}$ image slices were stacked to form 3D 1-channel volumetric images in the same manner. The rationality of separating $K^{trans}$ images from the 3-channel composite images is further addressed in Discussion.

The sizes of image patches were pre-defined to cover the areas from the regions of the small prostate lesions to the cross sections of the prostate gland; namely, the image patches were designed in four different sizes, including 42 × 42 × 1 × n, 48 × 48 × 3 × n, 64 × 64 × 3 × n, and 96 × 96 × 3 × n (height × width × depth × channel; the units of height and width in pixel and the unit of depth in the number of slices) in an associated 3D volumetric image, respectively. The number n, which is either 1 or 3, represents the composite channels. An image patch of 96 × 96 × 3 × n in size was first retrieved from the 3D volumetric image of a particular study, followed by generating the other three image patches of the smaller patch sizes with the same patch center as the image patch of 96 × 96 × 3 × n. The volumetric center of the image patch was the locus of a given clinical finding, recorded in the supplementary spreadsheet file of the PROSTATEx dataset, or was semi-randomly sampled from any coordinates that can retain the image patch of $96 \times 96 \times 3 \times n$ within the cropped 3D volumetric images (Fig. 1B). Semi-random sampling was used to combat the class imbalance by increasing the probability of sampling a clinical finding and its surrounding pixels by a factor of 10 when preceding image patch retrieval. For each study in the dataset, there were ~100 extracted image patches of single patch size, regarded as the inputs of the DenseNet-based model described below.

## Multiple-Stream Network Architecture

We propose a novel multiple-stream DenseNet architecture to extract hierarchical features from the image patch inputs automatically (Iandola *et al.*, 2014). Algorithm I sketches the pseudocode that provides a general understanding of the proposed neural network architecture layout. More specifically, four separate DenseNet streams built upon the four individual patch sizes were employed to reveal the feature maps of the respective source data possessing the particular patch sizes. As shown in Fig. 2A, the structure of a DenseNet stream transforms the input layer into a 3D convolutional layer and a successive maximum pooling layer, followed by three or four dense blocks, each of which is composed of a group of layers that are directly connected to all their subsequent layers in a feedforward fashion through concatenation. In a detailed manner, the single dense block is a composite function of three consecutive operations including a batch normalization, a ReLU activation, and a 3D $3 \times 3 \times 3$ convolution in our proposed network structure, in addition to a bottleneck layer consisting of a $1 \times 1 \times 1$ convolution, a batch normalization, and a rectified linear unit (ReLU) activation being introduced before the $3 \times 3 \times 3$ convolution to trim the number of input feature maps (Fig. 2B). Layers between two adjacent dense blocks are referred to as the transition layers, each of which, preceded by a batch normalization and a ReLU activation, also contains a $1 \times 1 \times 1$ convolutional layer followed by a $2 \times 2 \times 2$ (or $2 \times 2 \times 1$)

**Algorithm I** Four-Stream 3D DenseNet Architecture for Prostate Cancer Classification

**Data preprocessing**

Image scaling and cropping
Image intensity standardization
Extracting 4 different image patches

**Multi-stream DNN architecture**

***for*** image-patch size ***in*** pre-defined image-patch sizes ***do***
    ***for*** fold ***in*** 5-fold cross validation of the training cohort **do**
        train a stream architecture
- feedforward phase: feeding 3D MRI images of the image-patch size into 3D DenseNet backbone
- backpropagation learning phase: calculating the loss $L_f(\hat{y}_i, y_i)$, followed by updating weight values via $\partial L_f / \partial w$ in backpropagation

    ***end for***
    merge the trained model
***end for***
accumulate the separated, trained 3D DenseNet streams

**Ensemble learning**

train a second-level stacked generalization network
- feedforward phase: feeding the outputs from all the streams of **Multi-stream DNN architecture** into a 2-fully connected layer neural network
- backpropagation learning phase: calculating the loss $L_f(\hat{y}_i, y_i)$, followed by modifying weights via $\partial L_f / \partial w$ in backpropagation average pooling layer to archive the dimension reduction of feature maps (Fig. 2B). At the end of the last dense block, a global maximum pooling is appended to reduce the feature dimension and the output is fed into the succeeding softmax logistic classifier to estimate the probabilities for each class label in every individual DenseNet stream.

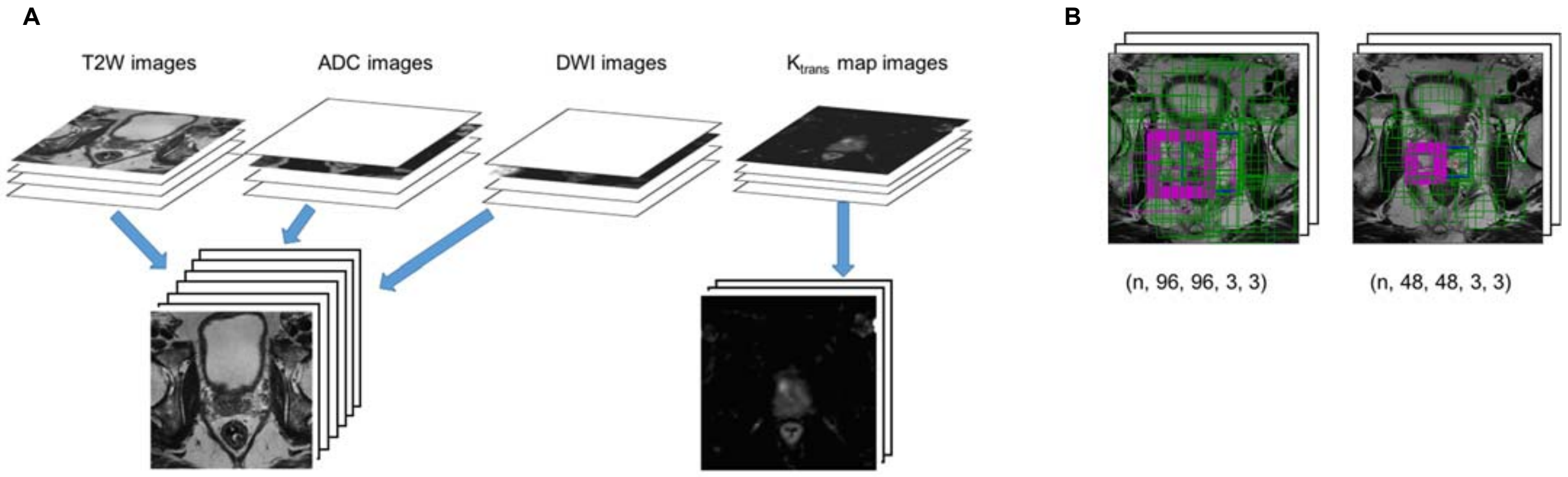

**Figure 1** Aligning MRI images retrieved from T2W, ADC, DWI, and $K^{trans}$ modalities according to the clinical lesions' coordinates, followed by image stacking. (A) T2W, ADC, and DWI images are concatenated to form the 3D 3-channel composite image, while the corresponding $K^{trans}$ image slices are stacked to generate the 3D 1-channel image. (B) The retrieved image patches with different dimensions are delineated in pink or green outlined squares.

**Cross-Validation Ensemble**

The image patches of a specified patch size in the training cohort were shuffled and split to implement 5-fold cross-validation; four of the five folds are used for model construction while the hold-out fold is allocated to model validation. After iterating through the folds, five individual trained models from the same architecture were generated and then accumulated to form a cross-validated model for a particular DenseNet stream. The same process was applied to the other three DenseNet streams so that the outputs of the individual DenseNet streams were collected as the inputs of the ensemble learning model described in the next paragraph.

**Stacked Generalization Ensemble**

Upon assembling the cross-validated DenseNet architectures, we implement a second-level meta-algorithm to establish the stacked generalization framework that optimizes the prediction accuracy (Ma *et al*., 2018), as shown in Fig. 2A. Specifically, these four individual cross-validated DenseNet models were embedded in the meta-machine learning model of a multi-headed neural network. Each of the cross-validated models was considered a separate input-head to the meta-stacking ensemble model. All the layers of the loaded DenseNet architecture were locked to prevent the weights from being modified when training the ensemble model. The outputs of the multimodal cross-validated DenseNet streams were concatenated to form a 20-element 1D array, referring to the integration of predicted probabilities from the corresponding four streams of 5-folded cross-validated models. The array was further mapped by an additional subset of two fully connected layers in the ensemble network to the final output, presenting the predicted probabilities of clinical significance in prostate cancer. In addition to presenting the prediction performance on the training cohort, we further validated the ensemble network by showing the result of the test cohort predictions.

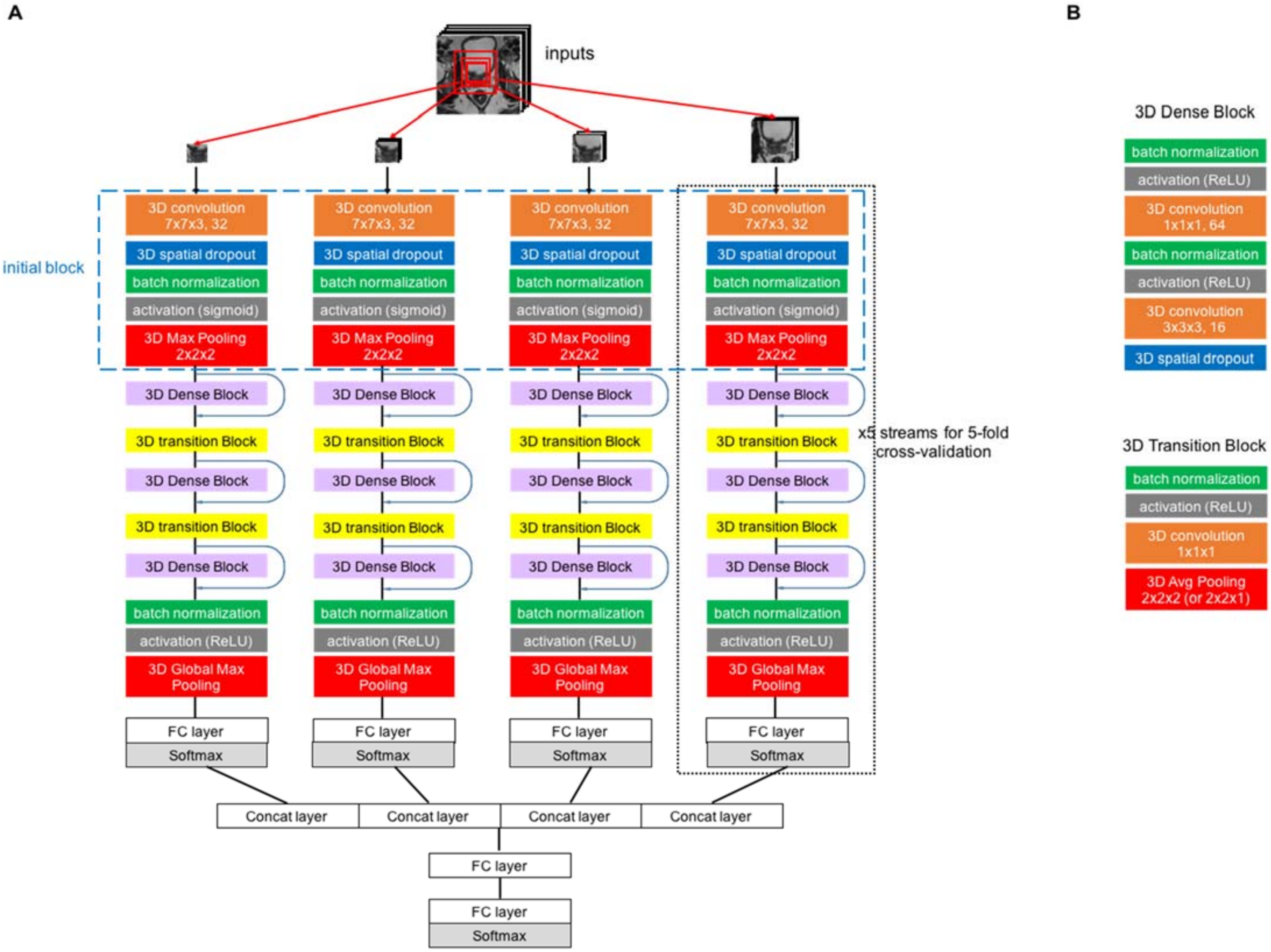

**Figure 2.** The framework of the proposed two-stage multimodal 3D DenseNet stream ensemble network. (A) The image patches of four different sizes are retrieved from the input images and fed into the specified DenseNet-based stream. All the four streams possess the same structure of the initial block, delineated in the blue dashed box. The black dotted box denotes the individual network stream replicates 5 times to fulfill the requirement of processing 5-fold cross-validation before entering the second stage of ensemble learning. (B) The corresponding dense block and transition block in the DenseNet-based backbone.

**Loss function and optimization for network update**

Typically, samples of clinical insignificance predominate over the subordinate samples of clinical significance in the medical image datasets, which results in the so-called imbalance classification problem in machine learning. The majority of easily classified samples in an unbalanced dataset are likely to either deliver useless learning signal that introduces inadequate training or overwhelm training that leads to model degeneracy (Krawczyk, 2016). To more effectively eliminate the impact of class bias, we adopt a novel loss function named focal loss function, which was designed originally to deal with an extreme imbalance between foreground and background classes during training for dense object detection (Lin *et al*., 2020).

$$FL(p_t) = -\alpha(1-p_t)^{\gamma}\log(p_t) \quad (1)$$

, where $\gamma$ is the adjustable focusing parameter and $p_t$ is defined as:

$$p_t = \begin{cases} p & if\ y = 1 \\ 1-p & if\ y = -1 \end{cases} \quad (2)$$

the estimated probability attributed to the class label $y$. Considering the implementation of a sigmoid function in the top layer for exporting the prediction in probability, $p_t = 1/(1+e^{-yX})$ , the focal loss can be expressed as

$$FL(p,y) = PL(X,y) = (1+e^{yX})^{-\gamma}\log(1+e^{-yX})$$
$$= (1+e^{X_t})^{-\gamma}\log(1+e^{-X_t}) \quad (3)$$

after defining $X_t = yX$ . Here $X$ is the output from the previous layer of the top layer.

The focal loss function replaces the conventional cross-entropy loss used in multiclass classification training with a dynamically scaled approach. As training proceeds, the scaling factor on samples of high confidence in predicting the correct class lessens over time, directing the model's attention towards samples with low probability for correct classification. Therefore, we anticipate that the focal loss function can reduce the degree of overfitting and improve network calibration.

Stochastic Gradient Descent with a Nesterov momentum of 0.9 was used as the optimizer in the present work. The learning rate was set as $2e^{-4}$, and the weight decay was activated with a value of $1e^{-5}$. The number of training epochs was set as 200, coupled with a batch size of 72. Early stopping was equipped to avoid overfitting once the model performance stopped improving on the validation set.

**Evaluation metrics**

The receiver operating characteristic (ROC) analysis is implemented as a figure of merit to report the performance of identifying malignant prostate lesions of clinical significance from clinical insignificances or benign lesions observed in the images. Each point on the ROC indicates a sensitivity-specificity pair associated with a particular decision threshold while sensitivity and specificity are defined as:

$$sensitivity = \frac{TP}{TP+FN},\ specificity = \frac{TN}{TN+FP}, \quad (4)$$

where TP, TN, FP, and FN denote the numbers of true positives, true negatives, false positives, and false negatives, respectively. The area under the ROC curve (AUC) is selected as the metric that evaluates a built model operating on the same image dataset in the challenge paradigm.

**Implementation**

Our model framework was developed mainly on top of Keras, a high-level neural network API using an open-source platform named TensorFlow as a backend. The model was trained and evaluated on a compute node with a dual Intel Xeon E5-2680v4 processor and two NVIDIA Tesla V100 GPU co-processors on the Midway High-Performance Computing (HPC) cluster at the University of Chicago Research Computing Center.

## Results

To investigate the performance of the proposed model during the training stage, we first compare the individual 3D DenseNet streams with the respective input image patch sizes of (42, 42, 1, 3), (48, 48, 3, 3), (64, 64, 3, 3), and (96, 96, 3, 3); each image patch consists of three channels (i.e., T2w, DWI, and ADC), with either one or three image slices. Fig. 3 presents the evolution of the loss values on the training and validation sets along the four 3D DenseNet streams. The L2 regularization applied on the model weights impacts the loss value, leading to lessen the robustness of presenting classification performance. Still, the downward trends of the loss curves towards the ends of training and validation processes reveal that the competitive performances of the stream models with the input sizes of (64, 64, 3, 3) and (96, 96, 3, 3) in Figs. 3C and 3D are slightly better than that of the stream model with the input size of (48, 48, 3, 3), shown in Fig. 3B. Further, all the three DenseNet stream models outperform the model with the input size of (42, 42, 1, 3) in Fig. 3A. The observation is also shown in Table I, consisting of the performances collected from all the stream models for each fold in the 5-fold cross-validation, of which the validation results were calculated from the ROC curves illustrated in Figure S1. The classification accuracy/ AUC in the training and validation stages ranges from 0.843 ± 0.006/ 0.920 ± 0.004 and 0.903 ± 0.014/ 0.920 ± 0.004 in average, respectively, for the stream model of the (42, 42, 1, 3) input dimensions up to 0.925 ± 0.003/ 0.978 ± 0.001 and 0.983 ± 0.001/ 0.978 ± 0.001 in average for the stream model with the input dimensions of (64, 64, 3, 3). While considering the clinical findings retrieved from the validation set that contains the clinical finding image patches and the semi-randomly sampled image patches, we observe that the accuracies and AUCs are comparable with or higher than those in the corresponding validation sets across the different stream models. This observation indicates no overfitting occurring in any of these four DenseNet stream models, and even the original clinical findings are validated alone.

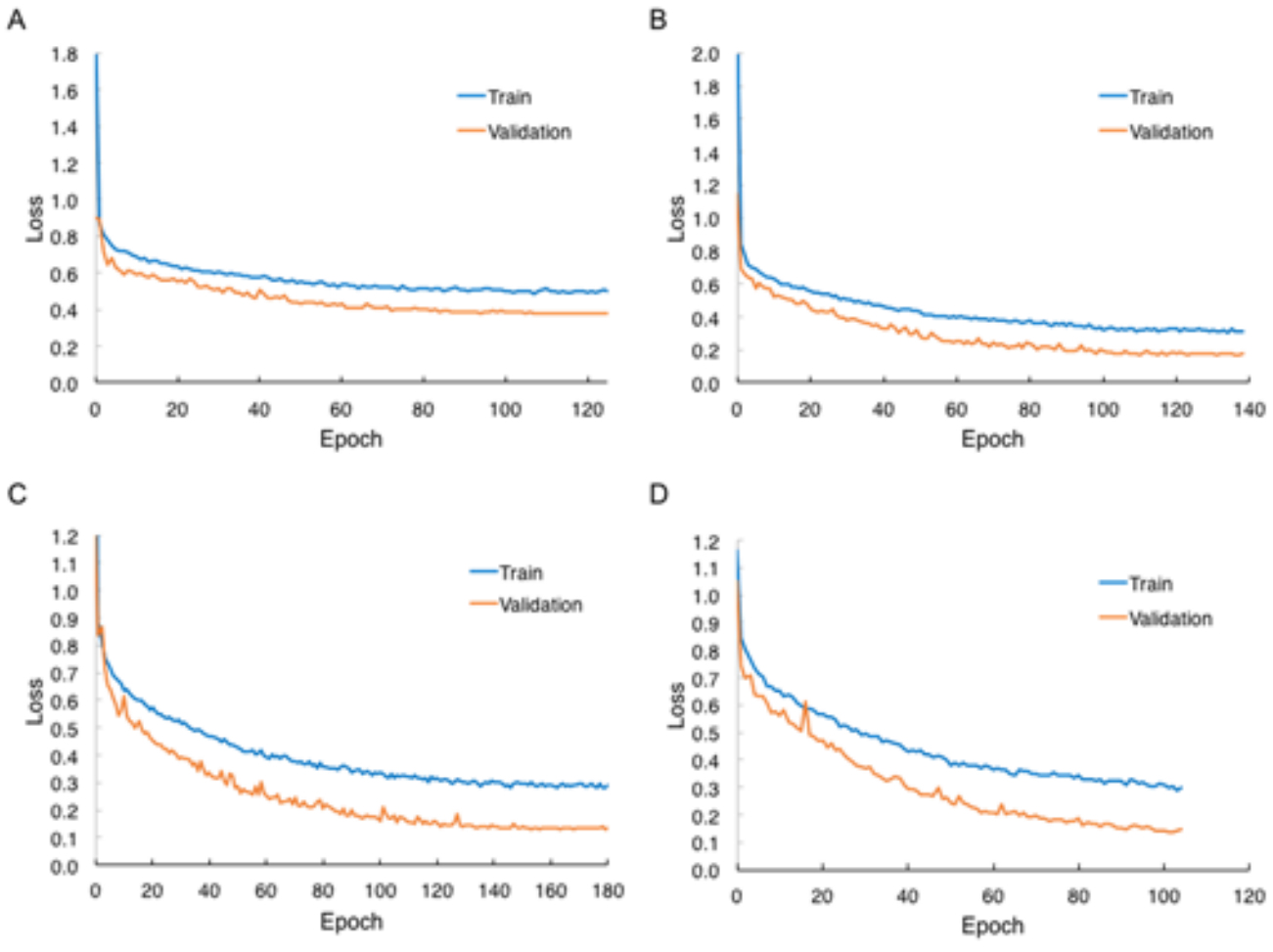

**Figure 3** Training and validation losses on the 3D DenseNet streams with three MRI modality channels in the training phase. The dimensions of image patch inputs are (A) (42, 42, 1, 3), (B) (48, 48, 3, 3), (C) (64, 64, 3, 3), and (D) (96, 96, 3, 3), respectively.

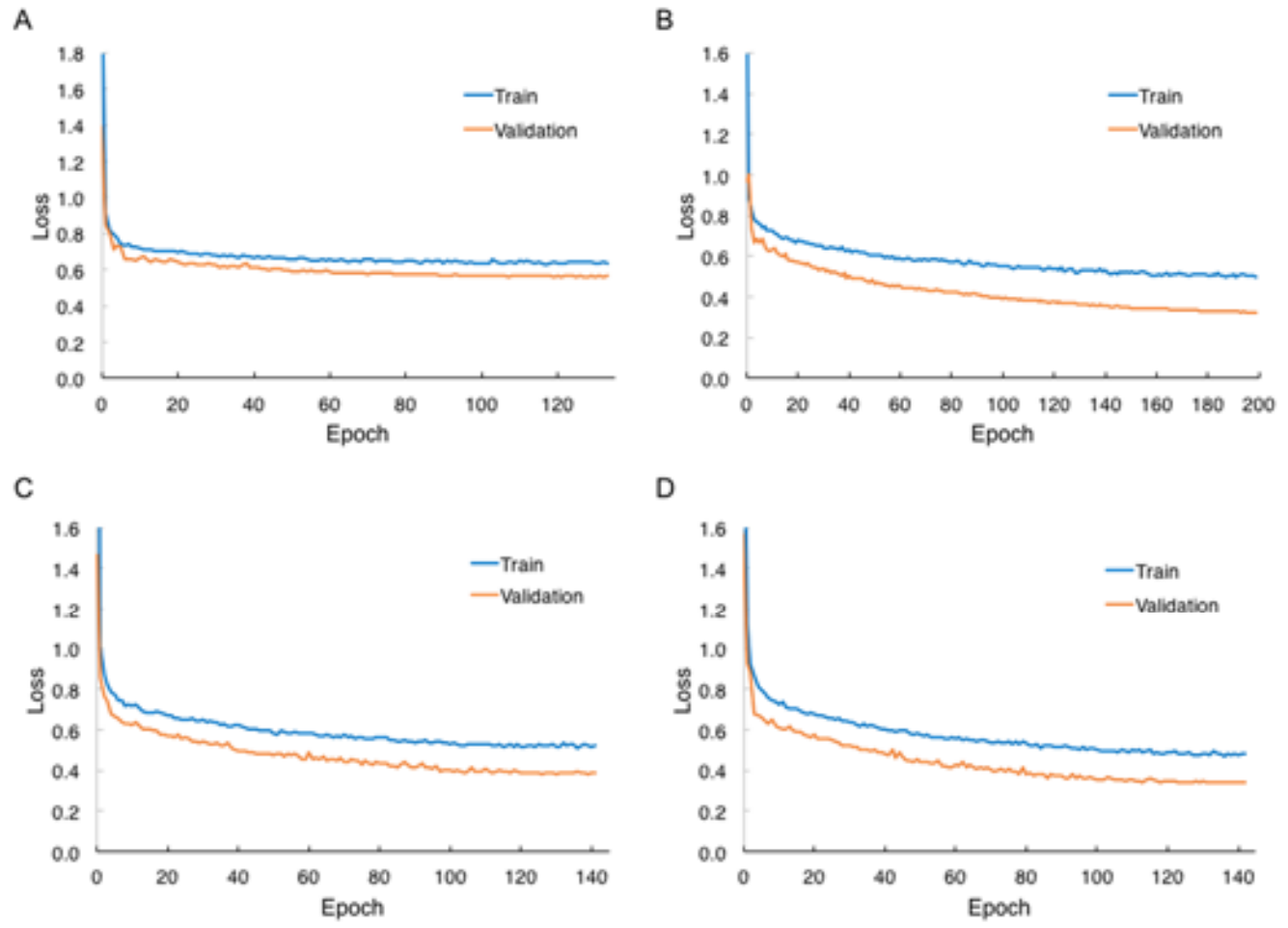

**Figure 4** Training and validation losses on the 3D DenseNet streams with single $K_{trans}$ channel in the training phase. The dimensions of image patch inputs are (A) (42, 42, 1, 1), (B) (48, 48, 3, 1), (C) (64, 64, 3, 1), and (D) (96, 96, 3, 1), respectively.

When considering $K^{trans}$ the only channel in the input image patches, the performance of the DenseNet stream models shown in Fig. 4 revealed the same trend as illustrated in Figure 3. The training and validation losses in the DenseNet stream model of the (42, 42, 1, 1) input matrices depict greater values than those shown in the other three DenseNet stream models. Table II also supports the observation in Fig. 4 that the classification accuracies/ AUCs on the training and validation regimes are 0.753 ± 0.004/ 0.843 ± 0.002 and 0.817 ± 0.004/ 0.843 ± 0.002 in average, respectively, for the stream model of the (42, 42, 1, 1) input dimensions. These values rise up to 0.852 ± 0.005/ 0.935 ± 0.003 and 0.930 ± 0.004/ 0.935 ± 0.003, respectively, for the stream model possessing the (96, 96, 3, 1) input image patches. In addition, the accuracy and AUC performance for the clinical finding retrievals within the validation set were comparable with or improved from those in the associated validation set, which is in accord with Table I.

Regarding prediction performance, the stacked generation ensemble delineated in Table III is better than any individual stream model presented in Table I for the metrics of accuracy and AUC both in the training and validation phases. Using the input patches of all the four MRI modalities, the ensemble framework reported the highest accuracy and

AUC values compared to using single or triple image modality channels. In addition, the performance in the models using the input patches of three channels is consistently superior to that in the models using the input patches of single channel regardless of the implementation of the stacked generation ensemble.

In the test cohort, the 142 subject studies contained 208 prostate lesions accompanied by their corresponding spatial coordinates but not clinical significance annotations. We implemented the proposed model to classify all the 208 clinical lesions as the likelihoods of being clinically significant, from which the result was submitted to the PROSTATEx organizer. When the ensemble framework cantaining the quadruple input channels of T2w, ADC, DWI, and $K^{trans}$ was used to predict the classes of prostate lesions, our proposed framework outperformed most of the methods submitted from the participating groups during the PROSTATEx Challenge as well as those attempts of exhibiting their strategies submitted after the Challenge. In addition, our approach shared the highest AUC value of 0.87 reported by the top competitor (Armato *et al.*, 2018). Even the ensemble framework with the triple input channels (i.e., T2w, ADC, and DWI) was applied to the task, the resultant AUC value still achieved 0.85, better than the second-highest AUC value in the challenge. Similarly, the predictions of the ensemble framework adopting the single $K^{trans}$ channel resulted in an AUC value of 0.83, comparable with that of the second-highest AUC value from the competition.

**Table I Performance in patch-size dependent stream models built on three MRI modalities**

| *image patch size | training | | validation | | clinical findings in validation | |
|---|---|---|---|---|---|---|
| | accuracy | AUC | accuracy | AUC | accuracy | AUC |
| (42, 42, 1, 3) | 0.839 | 0.920 | 0.900 | 0.920 | 0.958 | 0.998 |
| | 0.836 | 0.914 | 0.890 | 0.914 | 0.921 | 0.971 |
| | 0.846 | 0.923 | 0.910 | 0.923 | 0.925 | 0.983 |
| | 0.850 | 0.925 | 0.923 | 0.925 | 0.910 | 0.956 |
| | 0.842 | 0.920 | 0.891 | 0.920 | 0.895 | 0.950 |
| (48, 48, 3, 3) | 0.914 | 0.973 | 0.978 | 0.973 | 0.991 | 0.995 |
| | 0.913 | 0.973 | 0.981 | 0.973 | 0.980 | 0.997 |
| | 0.924 | 0.974 | 0.984 | 0.974 | 0.976 | 0.999 |
| | 0.916 | 0.974 | 0.982 | 0.974 | 0.972 | 0.995 |
| | 0.926 | 0.975 | 0.980 | 0.975 | 0.956 | 0.982 |
| (64, 64, 3, 3) | 0.923 | 0.977 | 0.983 | 0.977 | 0.967 | 0.999 |
| | 0.929 | 0.980 | 0.984 | 0.980 | 0.991 | 1.000 |
| | 0.922 | 0.978 | 0.983 | 0.978 | 0.981 | 0.998 |
| | 0.924 | 0.979 | 0.981 | 0.979 | 1.000 | 1.000 |
| | 0.927 | 0.978 | 0.985 | 0.978 | 0.967 | 0.984 |
| (96, 96, 3, 3) | 0.921 | 0.977 | 0.981 | 0.977 | 0.953 | 0.990 |
| | 0.926 | 0.978 | 0.983 | 0.978 | 0.982 | 0.997 |
| | 0.914 | 0.976 | 0.979 | 0.976 | 0.974 | 0.997 |
| | 0.928 | 0.977 | 0.979 | 0.977 | 0.978 | 0.994 |
| | 0.924 | 0.978 | 0.979 | 0.978 | 0.920 | 0.963 |

1. Image patch size denotes a respective stream model that the input layer passes the image patch inputs with the corresponding dimensions.
2. Accuracy represents the classification accuracy determined by a threshold of 0.5
3. AUC refers to the area under the ROC curve and the AUCs in the validation sets were calculated from the ROC curves presented in Fig. S1.
4. *"Clinical findings in Validation" indicates "performance in the clinical findings retrieved from the validation sets".

**Table II Performance in patch-size dependent stream models built on $K^{trans}$ images**

| *image patch size | training | | validation | | clinical findings in validation | |
|---|---|---|---|---|---|---|
| | accuracy | AUC | accurac y | AUC | accuracy | AUC |
| (42, 42, 1, 1) | 0.750 | 0.842 | 0.816 | 0.842 | 0.762 | 0.883 |
| | 0.755 | 0.842 | 0.813 | 0.842 | 0.781 | 0.858 |
| | 0.756 | 0.846 | 0.824 | 0.846 | 0.879 | 0.917 |
| | 0.755 | 0.845 | 0.815 | 0.845 | 0.783 | 0.873 |
| | 0.747 | 0.841 | 0.815 | 0.841 | 0.836 | 0.899 |
| (48, 48, 3, 1) | 0.847 | 0.933 | 0.934 | 0.933 | 0.950 | 0.991 |
| | 0.861 | 0.940 | 0.958 | 0.940 | 0.845 | 0.943 |
| | 0.849 | 0.934 | 0.938 | 0.934 | 0.926 | 0.986 |
| | 0.847 | 0.930 | 0.928 | 0.930 | 0.931 | 0.994 |
| | 0.848 | 0.934 | 0.942 | 0.934 | 0.969 | 0.997 |
| (64, 64, 3, 1) | 0.833 | 0.918 | 0.922 | 0.918 | 0.870 | 0.949 |
| | 0.832 | 0.918 | 0.902 | 0.918 | 0.919 | 0.973 |
| | 0.848 | 0.928 | 0.925 | 0.928 | 0.968 | 0.979 |
| | 0.830 | 0.919 | 0.904 | 0.919 | 0.855 | 0.959 |
| | 0.841 | 0.921 | 0.916 | 0.922 | 0.852 | 0.970 |
| (96, 96, 3, 1) | 0.843 | 0.931 | 0.932 | 0.932 | 0.964 | 0.989 |
| | 0.855 | 0.934 | 0.937 | 0.935 | 0.951 | 1.000 |
| | 0.856 | 0.938 | 0.926 | 0.938 | 0.889 | 0.933 |
| | 0.851 | 0.933 | 0.927 | 0.934 | 0.915 | 0.953 |
| | 0.855 | 0.938 | 0.930 | 0.938 | 0.899 | 0.966 |

1. Image patch size denotes a respective stream model that the input layer passes the image patch inputs with the corresponding dimensions.
2. Accuracy represents the classification accuracy determined by a threshold of 0.5
3. AUC refers to the area under the ROC curve and the AUCs in the validation sets were calculated from the ROC curves presented in Fig. S2.
4. *"Clinical findings in Validation" indicates "performance in the clinical findings retrieved from the validation sets".

**Table III Performance in the stacked generalization framework**

| Image patch input channels | training | | validation | | clinical findings in Validation | |
|---|---|---|---|---|---|---|
| | accuracy | AUC | accuracy | AUC | accuracy | AUC |
| T2w, ADC, DWI | 0.974 | 0.992 | 0.988 | 0.992 | 0.974 | 0.995 |
| $K_{trans}$ | 0.931 | 0.963 | 0.961 | 0.963 | 0.903 | 0.980 |
| T2w, ADC, DWI, $K_{trans}$ | 0.984 | 0.995 | 0.990 | 0.995 | 0.983 | 0.995 |

1. Accuracy represents the classification accuracy determined by a threshold of 0.5
2. AUC refers to the area under the ROC curve and the validation AUCs of the ensemble models with the respective input channels (T2w, ADC, DWI) and $K^{trans}$ were calculated from the ROC curves presented in Fig. S3.
3. *"Clinical findings in Validation" indicates "performance in the clinical findings retrieved from the validation sets".

Figure 5 shows an example of prostate cancer prediction achieved by our proposed model on the validation set. The model exported a probability of 0.95 of giving a positive result when applied to a clinically significant lesion. On the contrary, the model provides a probability of 0.01 of receiving a negative diagnosis for clinical insignificance.

## Discussion

In the past few years, CAD has demonstrated merits of medical image interpretation in clinical research, including automatic classification of malignant prostate lesions. However, those CAD research reports are typically linked with the selection of dataset composition, *subtle lesion* appearance, ground truth definition, and evaluation metrics, all of which barricade inter-method/ algorithm comparison (Nishikawa *et al*., 1994; Nishikawa and Yarusso, 1998; Revesz *et al*., 1983). We selected the PROSTATEx challenge for the architecture development because it provides a platform for direct comparison, given that all the participating algorithms employed a common dataset and were evaluated by an identical metric.

There are 330 prostate lesions in the training cohort of the PROSTATEx challenge, of which 23% are considered clinically significant, while another 208 prostate lesions excluding clinical significance information are in the test cohort. After examining the data distribution in the training cohort, it is evident that the PROSTATEx challenge uses a dataset of unbalanced class memberships, posing a challenge for predictive modeling. The imbalance in class membership may lead to overfitting on negative log-likelihood during the training process (Mukhoti *et al*., 2020), which is positively associated with the miscalibration of neural networks. Cross-entropy loss, a popular loss function for optimizing classification models, potentially leads to such an occurrence because it attempts to improve the confidence of the correct predictions rather than correct the predicting incorrectness after a certain number of epochs/ iterations. In other words, it focuses on reducing the difference between the probability distribution of a list of potential outcomes and the ground truths over an entire mini-batch. Therefore, we introduced the focal loss function to emphasize the misclassified subject samples rather than the correctly classified samples. Because the PROSTATEx challenge dataset possesses a moderate degree of class imbalance, the values of the hyperparameters $\alpha$ and $\gamma$ in focal loss we selected are 0.5 and 1.5, respectively, upon grid search for hyperparameter tuning.

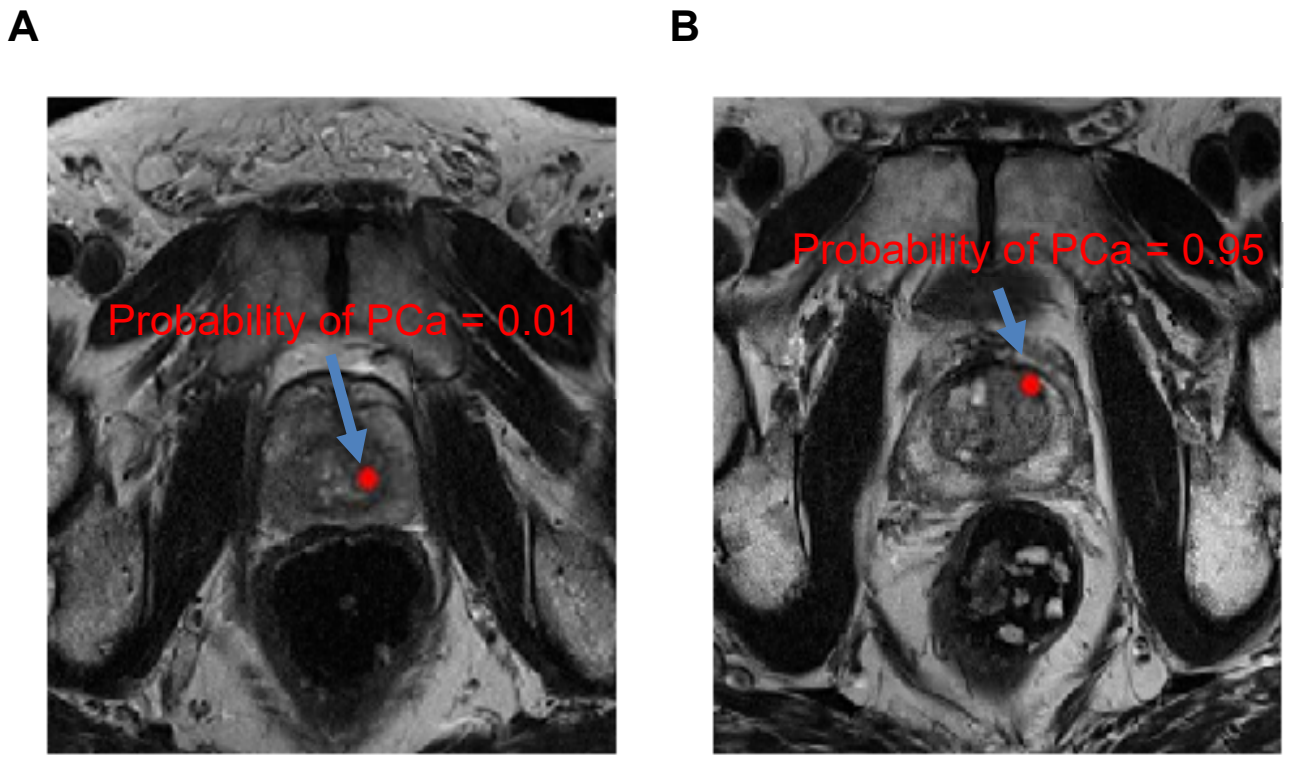

**Figure 5** Example of prostate cancer prediction on (A) a clinically insignificant lesion, and (B) a clinically significant lesion using the proposed two-stage quadruple-stream model

Similar to the other public medical image datasets, the PROSTATEx challenge is limited to the modest size of its dataset due to the difficulty of collecting a sufficient number of cases for training and testing, which may impact the statistical conclusion validity. To maximize the number of available samples, we retrieved the image patches from different loci within the same raw MRI images using the pre-defined patch dimensions. This process was conducted to populate the training set, leading to the overfitting reduction, described in the Method section. Furthermore, the proposed two-stage framework in the present work was designed by implementing quadruple 3D DenseNet stream architecture with the corresponding image patch inputs, on which the stacked generalization framework was established for meta-learning. Our approach successfully diminishes the prediction variance in a neural network model by training the multiple stream model and assembling the predictions from the individual stream networks instead of building a single stream model (Caruana *et al*., 2006). In addition to reducing the prediction variance, this so-called ensemble learning obtains more accurate predictions than any single stream model, as shown in Tables I, II, and III.

Several techniques, including batch normalization, L2 regularization, and dropout, were embedded in the proposed architecture framework to lessen overfitting. In the meantime, the greater loss values in the training phase than in the validation phase were revealed in Figs. 3 and 4, which would result from the certain processes or methods during model training and validating, including the L2 regularization that was applied in the training phase but not in the validation or testing phase. Moreover, the mechanism of dropout is to randomly disable hidden units, resulting in the subsequent layers attempting to construct the answers based on incomplete information. Therefore, it is artificially harder for the network to provide the correct predictions in the training phase. In contrast, all the hidden units are available in the validation process and the network retains the full computational power that may lead to better performance in the validation phase. In addition, the training loss was continuously measured and accumulated during each epoch, while validation loss was measured after each epoch had been completed. In other words, the training loss was measured half an epoch on average ahead of the validation loss measured.

The stacked generation ensemble reported the AUC value of 0.995 in the clinical prostate lesions of the validation set, while the AUC in the test cohort received from the organizer was 0.870. The decline in AUC values from the validation in the training cohort to the test cohort is mostly caused by sample selection biases, although several regularization techniques have been implemented. This phenomenon is also observed in Reference (Liu *et al.*, 2017). It is worth noting that the distribution of image intensity values in the training cohort is different from that of the test cohort, referring to the heterogeneity of image intensity between these two cohorts. The different distributions impact generalization and prediction performance as well as influence on predicting new cases not belonging to the PROSTATEx dataset.

There are a few advantages when adopting our proposed architecture framework. In contrast to the methods that also used the PROSTATEx challenge dataset, our approach does not need sophisticated processes for image registration or use 3rd party software packages for input image resampling and rescaling. Thus, images from the selected MRI modalities are automatically aligned upon the coordinates of the suspicious lesions, determined by the raters using the minimum labor cost. In addition, the meta-learning on the second level of our framework has the option of assembling the cross-validated DenseNet architectures with a single MRI channel ($K^{trans}$), three MRI channels (T2w, ADC, DWI), or all of the four MRI channels (T2w, ADC, DWI, $K^{trans}$). This strategy was enlightened after observing that concatenating the $K^{trans}$ image channel into the other three image modalities is not necessary to boost prediction performance significantly. In other words, stacking the $K^{trans}$ slices in the 3-channel composite images requires extra effort and receives the limited improvement. Therefore, the framework provides a specific option of selecting different combinations of MRI modalities if $K^{trans}$ images, for example, are unavailable or only provide insignificant improvement in the new cases.

This study aims to develop a novel neural network framework trained by a public annotated mpMRI dataset, through which our proposed method can fairly compare with the other reported CAD approaches for prostate cancer diagnosis. Undoubtedly, the modest size of the PROSTATEx dataset restricts generalization in the trained models, referring to the fundamental challenge in the collection of clinical medical images, including MRI. Further, the image intensity variations in MRI may result from any one or multiple causes, such as different MRI scanners and magnetic field strengths, different pulse sequence and parameter settings, different signal-to-noise ratios, and different operators. The image artifacts are frequently seen in MRI mainly due to magnetic field inhomogeneity, motion artifacts, or RF pulse and gradient pulse imperfection. Through comparing MRI images acquired in the prostate gland with those in the other organs such as brain, it is observed that the quality of prostate MRI is highly affected by subject movement that introduces different degrees of image displacement, image distortion, and motion artifacts among the different MRI modalities in each study case. Therefore, the necessity of image registration in medical imaging for training a deep learning model is an open-ended question and difficult to be generalized, which requires further study.

## Conclusion

We propose an efficient two-stage deep learning architecture using patchwise training for prostate cancer classification. All the steps of image preprocessing and data augmentation executed without any third-party software packages are integrated into the proposed framework before performing the end-to-end training. When assessing the test cohort predictions, the proposed model outperformed most of the submitted methods and shared the highest AUC value of 0.87 in the PROSTATEx Challenge (Armato *et al.*, 2018). The results reflect the promise of computer-aided diagnosis and neural networks in assisting medical image interpretation for prostate cancer detection and reducing unnecessary biopsies. Because the model was designed and trained upon the image patch inputs, it easily turns out to be a user-oriented cancerous detection, through which the user can point any locus to inquire its corresponding clinical significance, In the next steps, it is necessary to properly integrate future data into the training refinement for the purpose of model generalization

## Acknowledgment

The authors gratefully acknowledge funding support from Intel and Lenovo, USA. This work was completed in part with resources provided by the University of Chicago's Research Computing Center.

## Supplementary material

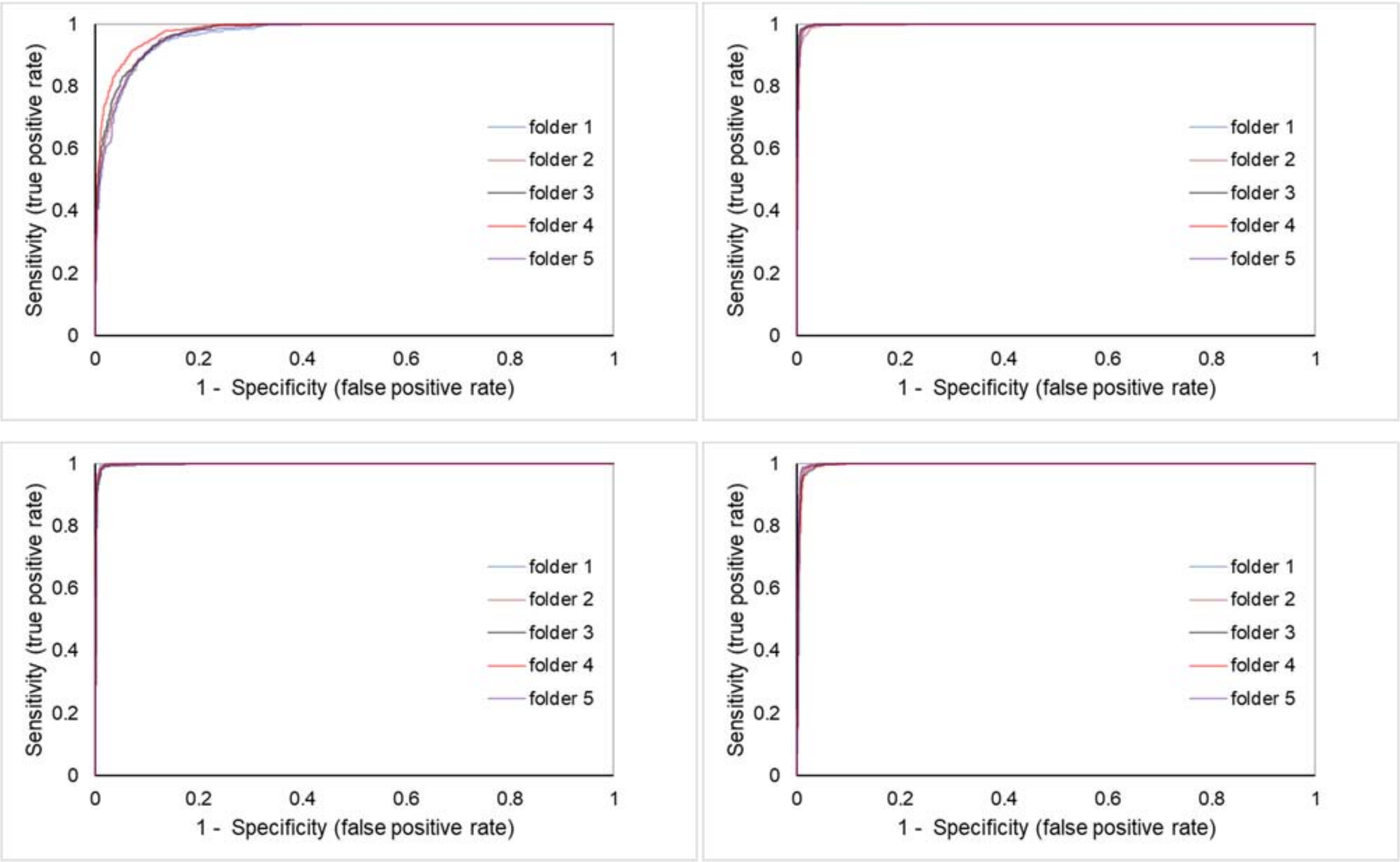

**Figure S1** ROC curves for evaluating the performance in classification of prostate lesions in PROSTATEx dataset into clinical significance and clinical insignificance/ benign categories, as described in the section of Evaluation Metrics. Classification was performed in (A) the (42, 42, 1, 3) stream model, (B) the (48, 48, 3, 3) stream model, (c) the (64, 64, 3, 3) stream model, and (D) the (96, 96, 3, 3) stream model, respectively, for each fold in the 5-fold cross-validation. The corresponding AUCs were presented on the column of Validation in Table I.

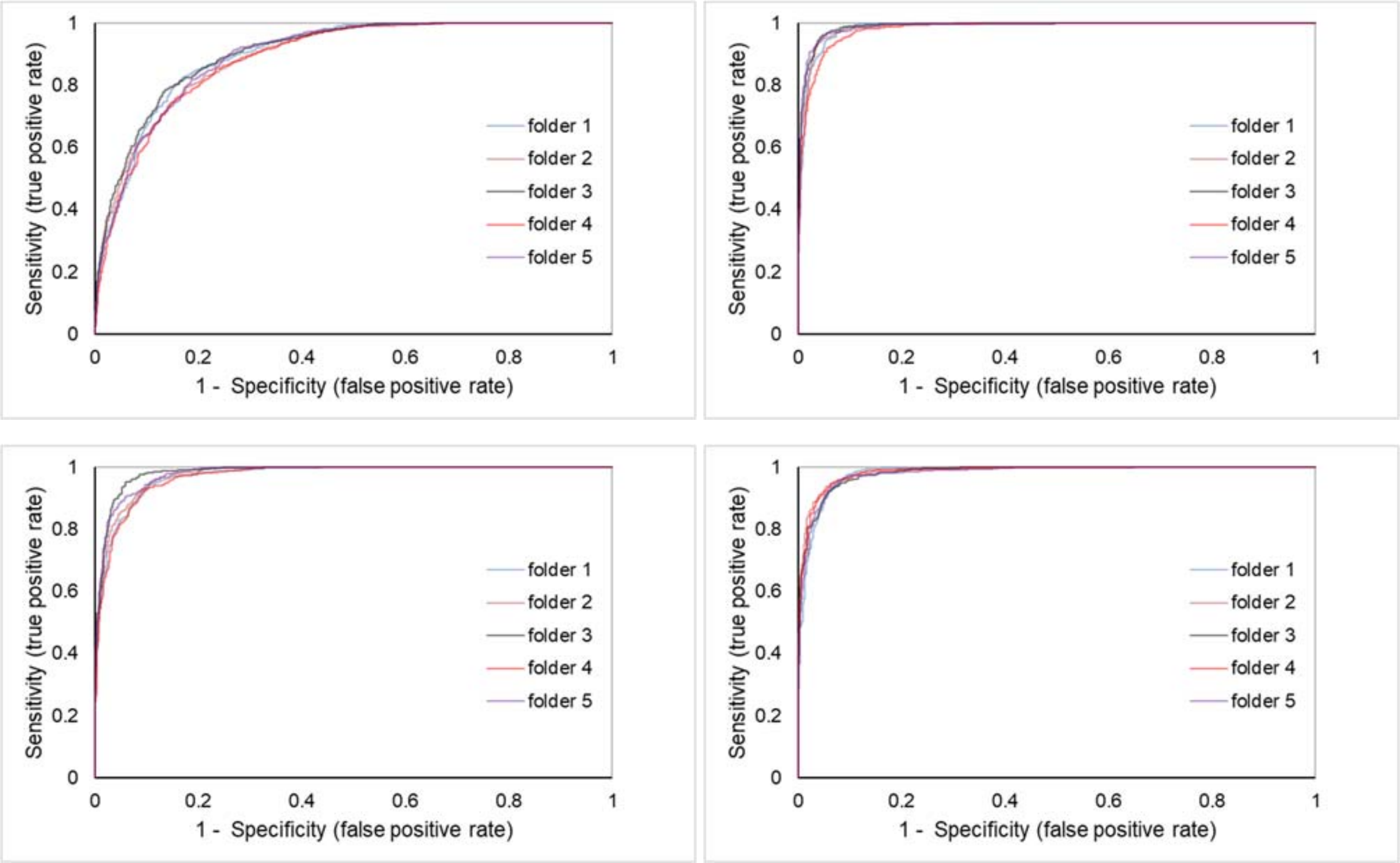

**Figure S2** ROC curves for evaluating the performance in classification of prostate lesions in PROSTATEx dataset into clinical significance and clinical insignificance/ benign categories. Classification was performed in the stream models upon the solo image modality, $K^{trans}$, which are (A) the (42, 42, 1, 1) stream model, (B) the (48, 48, 3, 1) stream model, (c) the (64, 64, 3, 1) stream model, and (D) the (96, 96, 3, 1) stream model, respectively, for each fold in the 5-fold cross-validation. The corresponding AUCs were presented on the column of Validation in Table II.

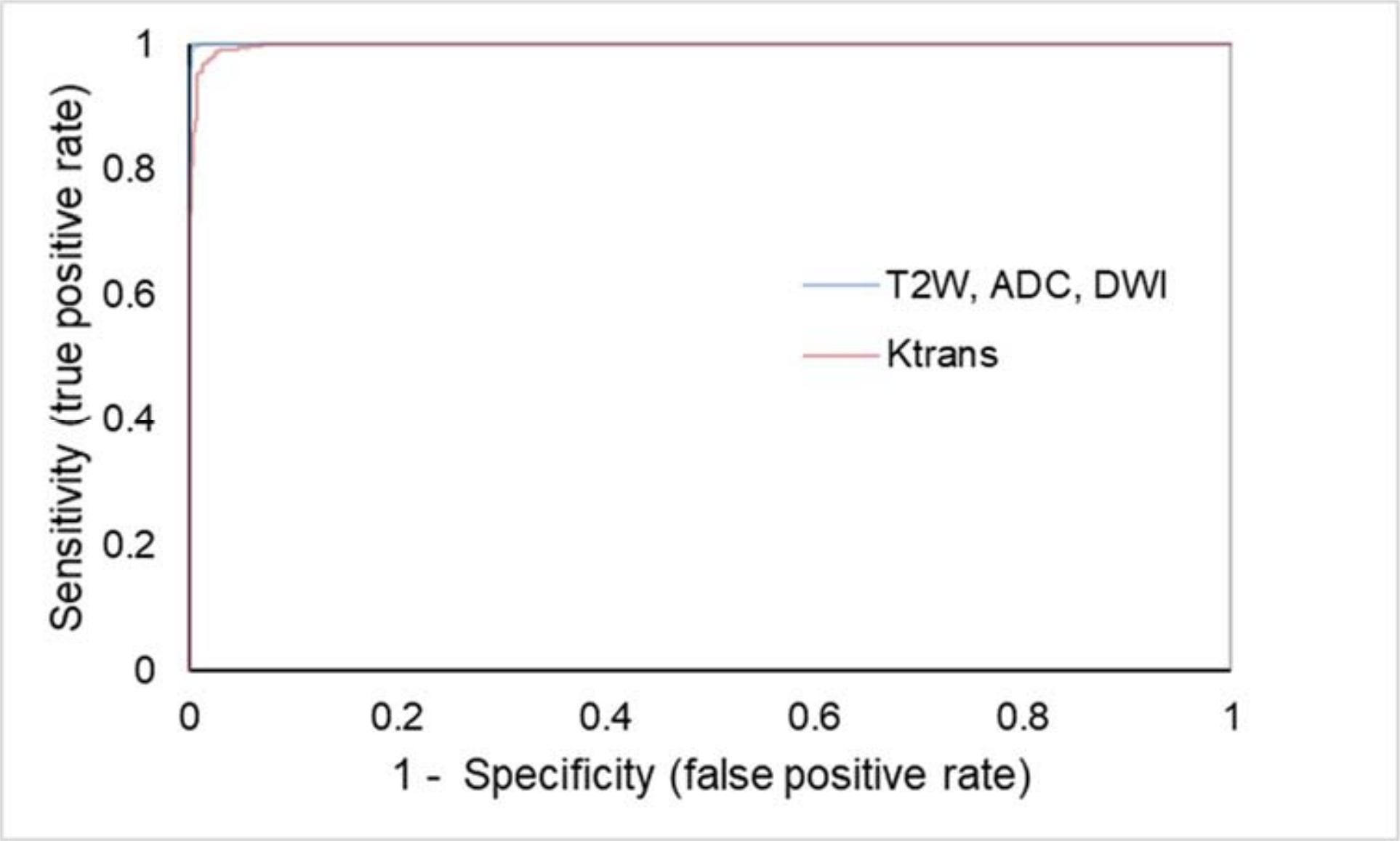

**Figure S3** ROC curves for evaluating the performance in classification of prostate lesions in PROSTATEx dataset into clinical significance and clinical insignificance/ benign categories. Classification was performed in the stacked generalization framework upon either the three stacked image modalities (T2w, ADC, and DWI) or the solo image modality, $K^{trans}$. The corresponding AUCs were presented on the validation column in Table III.